\documentclass[conference]{IEEEtran}
\IEEEoverridecommandlockouts

\usepackage[utf8]{inputenc}
\usepackage{acronym}
\usepackage{comment}
\usepackage[monochrome]{color}
\usepackage{amsmath}
\usepackage{algorithm}
\usepackage{algpseudocode}
\acrodef{sn}[SN]{satellite network}
\acrodef{an}[AN]{aerial network}
\acrodef{gn}[GN]{ground network}
\acrodef{rf}[RF]{radio frequency}
\acrodef{ris}[RIS]{reconfigurable intelligent surface}
\acrodef{vlc}[VLC]{visible light communications}
\acrodef{re}[RE]{reflective element}
\acrodef{geo}[GEO]{geostationary orbit}
\acrodef{meo}[MEO]{medium earth orbit}
\acrodef{leo}[LEO]{low earth orbit}
\acrodef{hap}[HAP]{high altitude platform}
\acrodef{lap}[LAP]{low altitude platform}
\acrodef{vn}[VN]{virtual network}
\acrodef{uav}[UAV]{unmanned aerial vehicle}
\acrodef{3d}[3D]{three-dimensional}
\acrodef{fso}[FSO]{free-space optical}
\acrodef{owc}[OWC]{optical Wireless Communications}
\acrodef{bs}[BS]{base station}
\acrodef{ber}[BER]{bit error rate}
\acrodef{thz}[THz]{terahertz}
\acrodef{mmwave}[mmWave]{millimeter wave}
\acrodef{ris}[RIS]{reconfigurable intelligent surface}
\acrodef{mimo}[MIMO]{multiple-input multiple-output}
\acrodef{los}[LoS]{line-of-sight}
\acrodef{nlos}[NLoS]{Non-LoS}
\acrodef{ir}[IR]{infrared}
\acrodef{uv}[UV]{ultraviolet}
\acrodef{ioe}[IoE]{Internet-of-Everything}
\acrodef{occ}[OCC]{optical camera communication}
\acrodef{lifi}[LiFi]{light fidelity}
\acrodef{led}[LED]{light-emitting diode}
\acrodef{ml}[ML]{machine learning}
\acrodef{qos}[QoS]{quality-of-service}
\acrodef{sagin}[SAGIN]{space-air-ground integrated network}
\acrodef{csi}[CSI]{channel state information}
\acrodef{em}[EM]{electromagnetic}
\acrodef{gan}[GAN]{generative adversarial network}
\acrodef{snr}[SNR]{signal-to-noise ratio}
\acrodef{sinr}[SINR]{signal-to-interference-plus-noise ratio}
\acrodef{e2e}[E2E]{end-to-end}
\acrodef{5g}[5G]{fifth generation}
\acrodef{mec}[MEC]{mobile edge computing}
\acrodef{mems}[MEMS]{microelectromechanical systems}
\acrodef{6g}[6G]{sixth generation}
\acrodef{b5g}[B5G]{beyond 5G}
\acrodef{tbps}[Tbps]{Tera-bit-per-second}
\usepackage{url}
\usepackage[pdftex]{graphicx}
\usepackage[
backend=biber,
style=ieee,
sorting=none
]{biblatex}
\usepackage{algorithm,algpseudocode}

\makeatletter
\newcommand\fs@spaceruled{\def\@fs@cfont{\bfseries}\let\@fs@capt\floatc@ruled
  \def\@fs@pre{\vspace{0.4\baselineskip}\hrule height.5pt depth0pt \kern2pt}%
  \def\@fs@post{\kern2pt\hrule\relax}%
  \def\@fs@mid{\kern2pt\hrule\kern2pt}%
  \let\@fs@iftopcapt\iftrue}
\makeatother

\title{Towards enabling reliable immersive teleoperation through Digital Twin: A UAV command and control use case}

\providecommand{\keywords}[1]
{
  \small	
  \textbf{\textit{Keywords---}} #1
}
\begin{document}

\author{
\IEEEauthorblockN{Nassim Sehad\textsuperscript{1}, Xinyi Tu\textsuperscript{2}, Akash Rajasekaran\textsuperscript{1}, Hamed Hellaoui\textsuperscript{1}, Riku Jäntti\textsuperscript{1}, and Mérouane Debbah\textsuperscript{3}}
\IEEEauthorblockA{\textsuperscript{1}\textit{Aalto University, School of Electrical Engineering,  Information and Communications Engineering, Espoo, Finland.}}
\IEEEauthorblockA{\textsuperscript{2}\textit{Aalto University, School of Engineering, Mechanical Engineering, Espoo, Finland.}}
\IEEEauthorblockA{\textsuperscript{3}\textit{Khalifa University of Science and Technology, Abu Dhabi, United Arab Emirates.}}
\IEEEauthorblockA{\textit{Emails: \{nassim.sehad, xinyi.tu, akash.rajasekaran, hamed.hallaoui, riku.jantti\}@aalto.fi, merouane.debbah@ku.ac.ae}}

\vspace{-0.8cm}
}

\vspace{-0.5cm}
\maketitle

\begin{abstract}
This paper addresses the challenging problem of enabling reliable immersive teleoperation in scenarios where an Unmanned Aerial Vehicle (UAV) is remotely controlled by an operator via a cellular network. Such scenarios can be quite critical particularly when the UAV lacks advanced equipment (e.g., Lidar-based auto stop) or when the network is subject to some performance constraints (e.g., delay). 
To tackle these challenges, we propose a novel architecture leveraging Digital Twin (DT) technology to create a virtual representation of the physical environment. This virtual environment accurately mirrors the physical world, accounting for 3D surroundings, weather constraints, and network limitations. 
To enhance teleoperation, the UAV in the virtual environment is equipped with advanced features that may be absent in the real UAV.
Furthermore, the proposed architecture introduces an intelligent logic that utilizes information from both virtual and physical environments to approve, deny, or correct actions initiated by the UAV operator. This anticipatory approach helps to mitigate potential risks. 
Through a series of field trials, we demonstrate the effectiveness of the proposed architecture in significantly improving the reliability of UAV teleoperation.



\end{abstract} 
\vspace{-0.18cm}
\section{Introduction} 

In recent times, the fusion of teleoperation and Internet of Things offers an enhanced paradigm for controlling Unmanned Aerial Vehicles (UAVs). At the nexus of this evolution are Extended Reality (XR) technologies, including Virtual Reality (VR), Augmented Reality (AR), and Mixed Reality (MR), which gained considerable attention thanks to their ability to provide a more intuitive and immersive control experience. Another pivotal advancement that complements this experience is haptic feedback technology that delivers tactile information to operators \cite{malik2020effect}.
\cite{hoppe2018vrhapticdrones} proposed VRHapticDrones, a system providing haptic feedback in VR without wearable devices, although limited by collision risks and trajectory planning complexities. Similarly, \cite{abtahi2019beyond} introduced HoverHaptics, autonomous quadcopters serving as haptic devices in VR. This work addressed challenges like speed limitations, control accuracy, and safety concerns using display techniques and collision avoidance strategies.

Despite these advancements, the use of UAVs for teleoperation is still facing challenges such as Network Latency (NL), environmental disturbances, and long-distance operation, which can lead to an unstable control of the remote system. 
This becomes more complex as UAVs could lack advanced equipment, such as Lidar-based auto stop.
 Furthermore, novel user interfaces, such as Brain-Computer Interfaces (BCIs) combined with XR for Human-Computer Interaction (HCI), have been shown to introduce significant reaction delays exceeding 400 ms \cite{kim2021p300}\cite{9951155}\cite{10071541}. However, these studies do not explore methods for adapting to network challenges or establishing autonomous control mechanisms in cases of incidents caused by high delays.
To address these challenges, we propose a novel architecture in this paper that enables reliable immersive teleoperation through the use of Digital Twins (DTs). The DT creates a virtual representation of a physical UAV, rendered in the VR environment. 

The main contributions of this paper are as follows:
\begin{enumerate}
\item Proposing a novel architecture for DT-enabled UAV immersive teleoperation in a VR environment, incorporating detailed 3D surroundings, weather, and network constraints;
\item  Enhancing the virtual UAV with advanced features absent in the real UAV, improving teleoperation reliability;
\item Introducing an intelligent decision-making logic that leverages information from both the virtual and physical environments to approve, deny, or modify actions initiated by the UAV operator, thereby anticipating and mitigating potential risks;
\item Implementing the proposed architecture and validating its effectiveness through a series of field trials, demonstrating its practical applicability.
\end{enumerate}

The remainder of this paper is structured as follows: Section~\ref{sec:DT} presents an overview of DT technology, discussing its various challenges when applied to teleoperation. Section~\ref{sec:scenario} introduces a motivating use case scenario that illustrates the benefits of employing DTs to enhance reliability in immersive teleoperation for UAV operators. In Section~\ref{sec:architecture}, we describe our proposed architecture for DT-enabled UAV teleoperation. Section~\ref{sec:results} provides a summary of the field trials conducted and the results obtained. Finally, Section~\ref{sec:conclusion} concludes the paper and highlights potential avenues for future research.

\section{Digital twins for teleoperation} 
\label{sec:DT}

\subsection{Overview}

DTs are digital replicas of physical assets, processes, and systems that can be used for simulation, analysis, and optimization in various fields including manufacturing, healthcare, and transportation \cite{liu2021review}. 

In the context of teleoperation, DTs have shown great potential to improve the efficiency and safety of remote systems \cite{rosen2015importance}. 
DTs can be used to monitor the health and performance of a remote system in real-time, enabling proactive maintenance and minimizing downtime. This is particularly important for UAV teleoperation, where any downtime can cause significant delays and disruptions. DTs can also be used to identify potential issues before they become critical, allowing for early intervention and prevention of more severe problems.
By simulating the behavior of a UAV in different environments, DTs can identify potential hazards and risks before a UAV is deployed. This allows better risk assessment and mitigation strategies, reducing the likelihood of accidents or incidents.
\vspace{-0.2cm}
\subsection{Challenges}
Despite the promising potential of DTs in teleoperation, there are still several challenges and research gaps that need to be addressed. 
One of the main challenges is the availability and quality of data. DTs require a large amount of data to accurately simulate the behavior of a remote system \cite{tu2021mixed}. However, the data is often fragmented and inconsistent. This can result in inaccuracies and errors of DTs, which further lead to suboptimal performance or even failures in real-world operations. Therefore, it is crucial to ensure that the data used for developing DTs is accurate, reliable, and up-to-date. This requires access to a variety of data sources, which, for UAV teleoperation, include sensors, telemetry data, and environment data. 

Another challenge is the complexity and flexibility of the models used to create DTs \cite{tao2022digital}. Models should be accurate to ensure DTs reliably reflect the behaviors of the remote system. Models should also be flexible enough to account for dynamic environments, where UAV teleoperation faces significant challenges such as changing weather conditions, interference from other wireless systems, and safety concerns for people and property. This requires sophisticated modeling techniques to capture the complex interactions between a UAV and its environment. 

In addition, the integration of DTs with other systems and processes in the teleoperation ecosystem can also be challenging, especially the synergies between DTs and XR \cite{10.3389/frvir.2023.1019080}, which is critical for immersive teleoperation. In the UAV case, data from DTs may need to be integrated with the UAV control system, the teleoperation platform, or its VR interface, to ensure seamless operation \cite{kikuchi2022future}. This requires a deep understanding of the technical requirements and constraints of each system, as well as the ability to design and implement integration solutions.

Finally, there are also technical challenges associated with the deployment of DTs for teleoperation in real-world environments \cite{sun2020reducing}. For instance, the computing resources required to run a DT in real-time may be significant, demanding high-performance computing infrastructure \cite{concannon2019quality}. 
This can lead to high power consumption, posing a challenge for the onboard computing of DTs in head-mounted devices (HMDs) \cite{alsamhi2022computing}.
In addition, the communication and networking infrastructure used to transmit data between DTs and the remote systems should be robust and reliable \cite{yang2022extended}, particularly in challenging environments for UAV remote control. Addressing these technical challenges requires a comprehensive understanding of the underlying technologies and infrastructures required to support DT-enabled UAV teleoperation. 

Overall, a standard architecture for integrating and developing DTs in UAV immersive teleoperation is still missing, which should cover data acquisition, modeling, simulation, and validation to ensure the accuracy and reliability of DT-enabled UAV command and control. 

\begin{figure*}[ht]
    \centering
    \includegraphics[width=\textwidth]{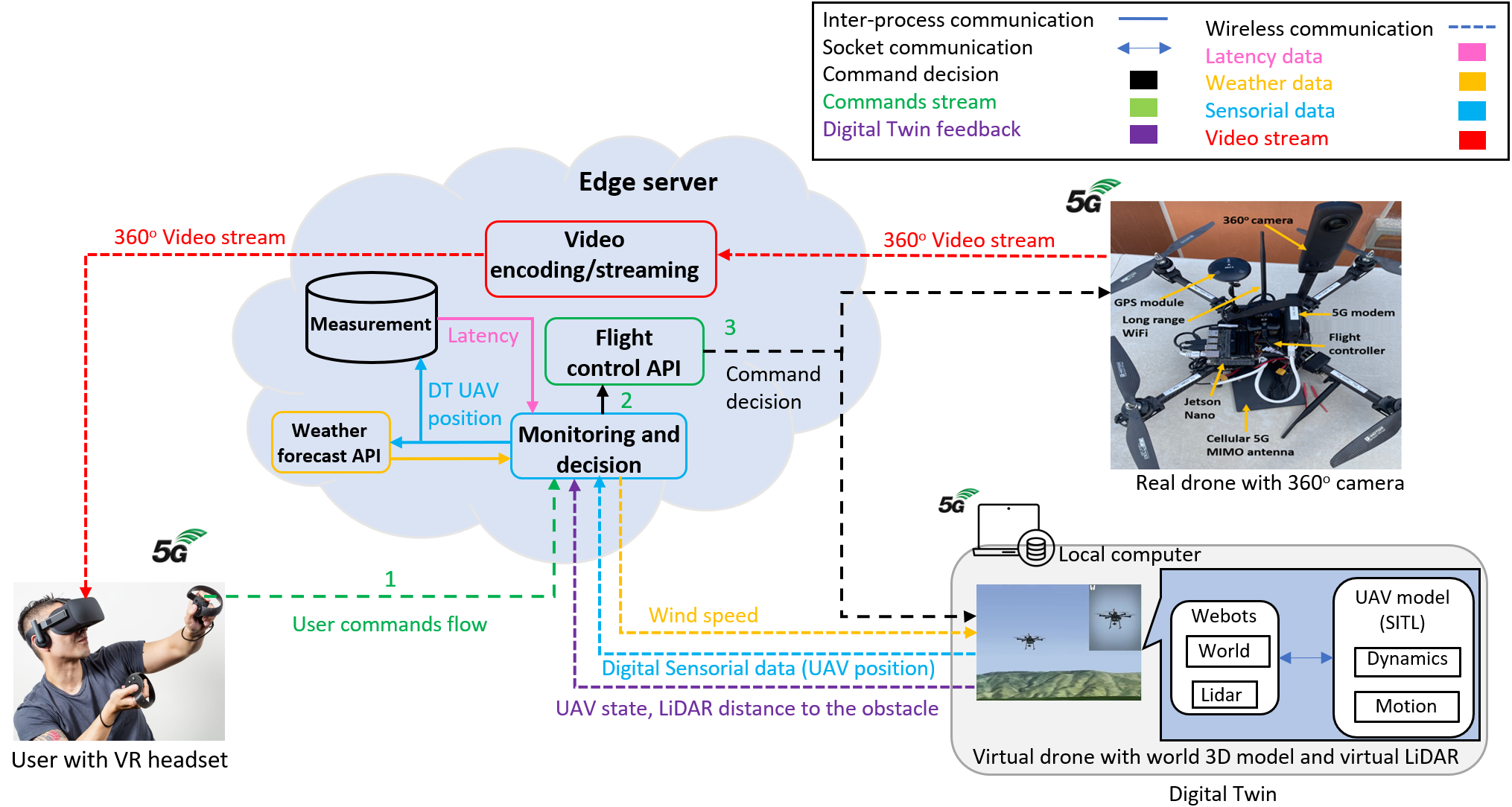}
    \caption{Our proposed architecture for digital twin-enabled UAV teleoperation.}
    \label{fig:my_label}
    \vspace{-0.5cm}
\end{figure*}
\section{A use case scenario}
\label{sec:scenario}
Thomas, an operational engineer for a power line inspection firm, pilots UAVs through challenging terrains like marshlands, forests, and mountains, using a 5G network. Using First Person View (FPV) mode and a HMD with joysticks, he captures images and videos of power lines.
Despite being a professional pilot, ensuring the safety of the operation is challenging. This is primarily due to long-distance operations, introducing significant delays, diminishing video quality and making safe piloting arduous. Misjudgments not only endanger the UAV but also the power grid.

To prevent the UAV from crashing, it is important to consider factors exacerbating the delays, primarily unstable connections between the UAV and Base Stations (BS) due to the mobility of the UAV, especially at high altitudes. Here, the reduced gains of the down-tilted BS antennas can lead to suboptimal handover and the selection of farther BSs \cite{hayat2019experimental}. Such issues reduce bandwidth and may degrade or interrupt video transmission.
Incorporating a DT to pre-emptively execute commands and monitor the state of the UAV can help mitigate these issues, allowing real-time adjustments to Thomas's piloting instructions.

\section{Our proposed architecture for digital twin-enabled UAV teleoperaton} 
\label{sec:architecture}

This section introduces a novel architecture for DT-enabled UAV teleoperation. The proposed architecture enables a user wearing an HMD to reliably control a remote UAV located anywhere in the world with 5G connectivity. 
Furthermore, the proposed architecture also incorporates intelligent logic to monitor different parameters from the virtual and the physical environments to approve/deny/correct actions initiated by the operator in a way to anticipate some risks.
The components of the proposed architecture are illustrated in Figure \ref{fig:my_label} and are explained in detail below.  
\vspace{-0.1cm}
\subsection{Real UAV}
The UAV is a quadcopter equipped with a 360$^{\circ}$ camera, an onboard computer, a flight controller, and a 5G modem. The onboard computer receives user commands via telemetry from the flight controller API module and sends captured video to the video module for encoding and streaming. Both modules are located on the edge server.
\vspace{-0.2cm}

\subsection{Virtual UAV/Digital twin}
The DT of the UAV and the world is a virtual representation that runs on a local computer situated near the edge server. Its primary responsibility is to simulate real-world physics and the UAV's behavior, thus providing a highly accurate and synchronized virtual model of the physical UAV. 

\begin{figure}[ht]
    \centering    \includegraphics[width=0.48\textwidth]{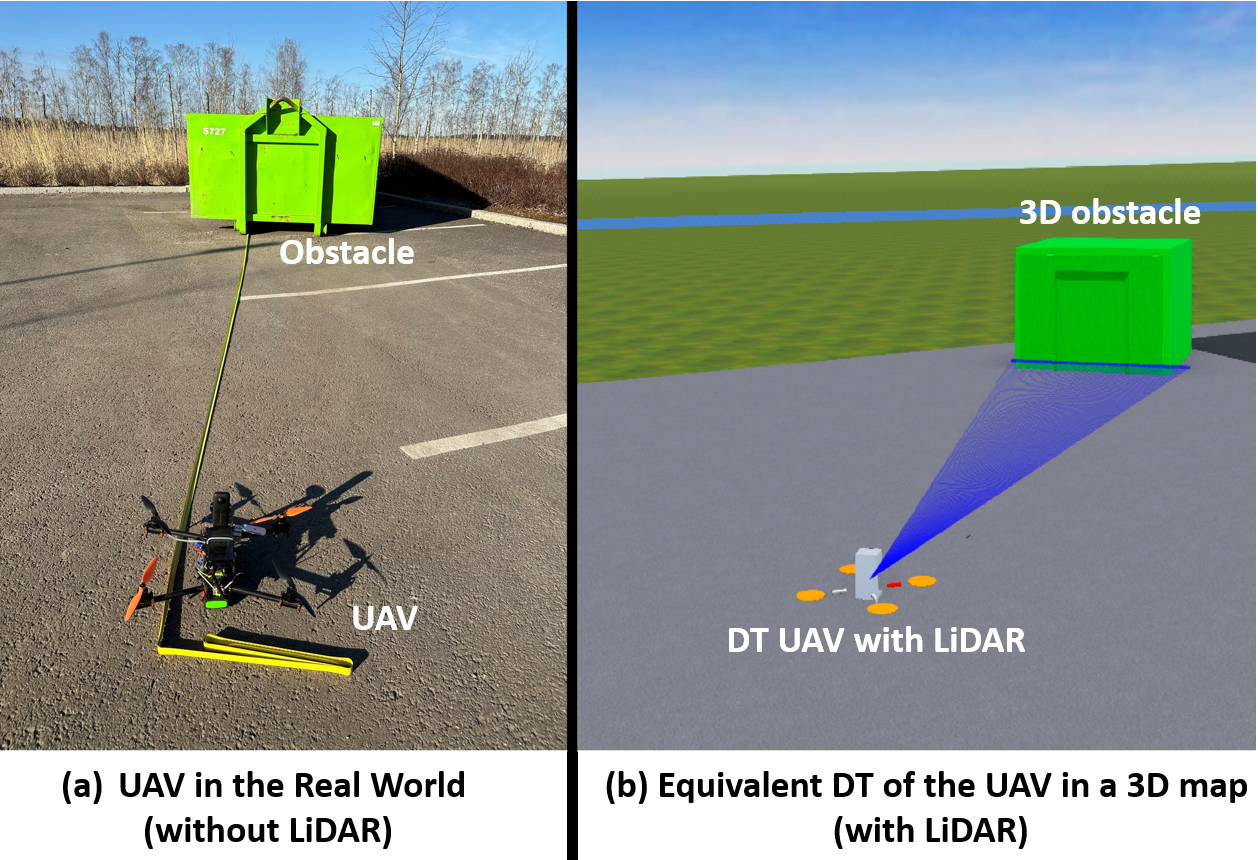}
    \caption{Comparison of the DT and the real-world UAV. }\vspace{-0.5cm}

    \label{fig:DT}
\end{figure}

The DT consists of two main components: a Webots simulation environment \cite{michel2004cyberbotics} and a UAV model from a Software In The Loop (SITL) simulator. The Webots environment includes the world simulation and a virtual LiDAR sensor that reports distance to facing objects, while the SITL simulator encompasses the dynamics and motion of the UAV. The Webots simulation and the UAV model are connected through socket communication. 
The world simulation in the DT replicates real-world physics within the virtual environment, taking into account various factors that a UAV may encounter in the field, such as weight, dimensions, and weather conditions, namely, wind speed, direction, and turbulence provided by the weather forecast API.
A virtual LiDAR sensor is integrated within the Webots simulation to accurately measure distances to nearby obstacles, enabling the virtual UAV to detect and avoid potential collisions in the simulated environment.
Figure \ref{fig:DT} illustrates the UAV in the real world (without LiDAR) alongside its DT equivalent within a 3D map, which is equipped with a LiDAR to prevent collisions with approaching obstacles. The figure also represents the experimental scenario utilized for validating the proposed architecture, as described in Section~\ref{sec:results} \textit{Results and Analysis}.

\subsection{VR user}

The VR user operates the UAV through a HMD with 5G connectivity, which displays the live omnidirectional video feed from the UAV's 360$^{\circ}$ camera in a virtual environment as if they were physically present at the UAV's location. The HMD is equipped with joysticks that allow the user to intuitively control the UAV's movements, including altitude, yaw, pitch, and roll. To ensure low latency and smooth control of the UAV, the HMD continuously sends consecutive commands through the MQ Telemetry Transport (MQTT) protocol to the monitoring and decision module on the edge server over a 5G network.

\subsection{Edge server}

The edge server is a crucial component of the DT-enabled UAV teleoperation system, as it serves as an intermediary between various entities, such as the real and virtual UAVs, the VR user, and external services. The edge server is designed to be modular, with five main modules handling distinct functionalities: video encoding/streaming, monitoring and decision, flight control API, weather forecast API, and a measurement database. These modules work in tandem to provide a seamless and reliable teleoperation experience.

The \textbf{video encoding/streaming module} is responsible for handling the 360$^{\circ}$ video stream received from the real UAV's camera. Upon receiving the video feed, it encodes and compresses the stream for efficient transmission, before sending it to the VR user's HMD. This ensures that the user has a real-time view of the UAV's surroundings, enabling them to make informed decisions while controlling the UAV.

The \textbf{monitoring and decision module} serves as the central hub for processing and managing data from various sources. It receives the user command flow from the VR user, UAV state and LiDAR distance data from the virtual UAV, wind speed data from the weather forecast API module, and sensorial data from the real UAV. Taking into account factors such as the current UAV state, environmental conditions, and potential obstacles, the module makes command decisions based on this information. It is important to note that the real UAV and its digital twin have identical starting positions for any given mission.
This module incorporates a logic allowing to approve/deny commands initiated by the operator. 
The decision to forward the user's command to the flight control API module is based on whether the condition in Equation~(\ref{eq:condition}) is met. This equation is provided as
\begin{equation}\label{eq:condition}
\small
LD \leq TH + EDL,
\end{equation}
\noindent
where the LiDAR Distance ($LD$) refers to the distance to a facing object reported by the DT UAV's LiDAR, and the ThresHold ($TH$) is a fixed predefined distance at which the UAV is programmed to stop.
$EDL$ stands for the Error Distance related to the Latency and is defined in Equation~(\ref{eq:error_distance}) as
\begin{equation}
\label{eq:error_distance}
\small
EDL = (CL + ENL) \times US,
\end{equation}

\noindent
where $CL$ stands for the average of a measured Command Latency between the DT and the real UAV which variates according to the Estimated Network Latency ($ENL$).
As for $US$, it refers to the UAV's Speed in the DT/Real-world at position coordinates (${U_x, U_y, U_z}$).

This allows Delay Compensation (DC) and collision avoidance following Algorithm \ref{alg:monitoring-decision}, facilitating effective decision-making.
\floatstyle{spaceruled}
\restylefloat{algorithm}
\begin{algorithm}[ht]
  \caption{DC-based Monitoring and Decision Algorithm}\label{alg:monitoring-decision}
  \begin{algorithmic}[ht]
\Require $\vec{U} = (U_x, U_y, U_z)$, $LD(t)$, $TH=1m$, $CL=146ms$
\Function{Monitoring\_Decision}{}
  \State $EDL \gets US \times [CL + \operatorname{GET\_ENL}(\vec{U})]$

   \If{$LD(t) \leq TH + EDL$ \textbf{and} $command\_direction = forward$}
    \State Send stop command to the flight control API
  \Else
    \State Forward command to the flight control API
  \EndIf
\EndFunction

\Function{get\_ENL}{$\vec{U}$}
  \State $closest\_NL \gets \text{null}$
  \State $min\_distance \gets \infty$
  \State $NL \gets \operatorname{read\_NL\_from\_file}("measurement")$
  \For {\textbf{each} $NL$ \textbf{in} $measurement$}
\State \parbox[t]{1cm}{$Dist  \gets  \operatorname{compute\_distance}(\vec{U}, NL.lat, NL.lon, NL.alt)$}

    \If{$Dist < min\_distance$}
      \State $ENL \gets NL$
      \State $min\_distance \gets distance$
    \EndIf
  \EndFor
  \State \Return $ENL$
\EndFunction
  \end{algorithmic}
\end{algorithm}

The \textbf{flight control API module} is responsible for relaying command decisions from the monitoring and decision module to both the real and virtual UAVs. By ensuring that the command decisions are executed on both UAVs, this module maintains synchronization between the real and DT UAVs. This allows for accurate testing and validation of command decisions in the virtual environment before applying them to the physical UAV, thus reducing the risk of accidents and improving operational efficiency.

The \textbf{weather forecast API module} communicates with an external weather service to obtain real-time wind speed data for the UAV's location. 
It receives  the UAV's position from the monitoring and decision module and sends the wind speed information in return. 
By considering the impact of wind on UAV performance, this module helps improve the accuracy and safety of UAV operations.

The \textbf{measurement module} maintains a database containing information about the average NL between the UAV and the edge server, based on the specific BS to which the UAV is connected. This database is constructed from measurements, as elaborated in Section~\ref{Measurement process} \textit{Measurement Process}. The database is indexed by the current cell ID  that the UAV is connected to and the UAV's position, which are obtained from the UAV's cellular modem and Global Positioning System (GPS) respectively. These indices are used to provide delay estimates for the communication between the UAV and the edge server.

\subsection{Measurement process}
\label{Measurement process}

\begin{figure}[ht]
    \centering \includegraphics[width=0.45\textwidth]{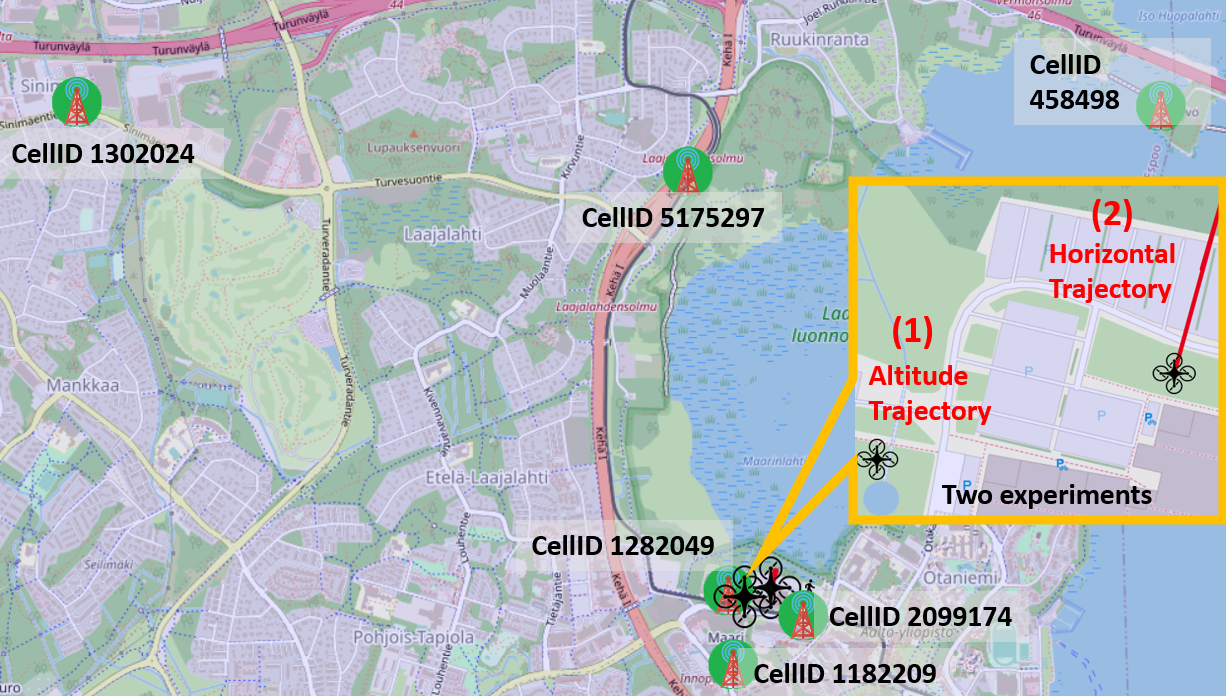}
    \caption{UAV flight paths along serving base station IDs and locations with (1) altitude trajectory and (2) horizontal trajectory. }
    \label{fig:paths}
\end{figure}

\begin{figure}[ht]
    \centering \includegraphics[width=0.48\textwidth]{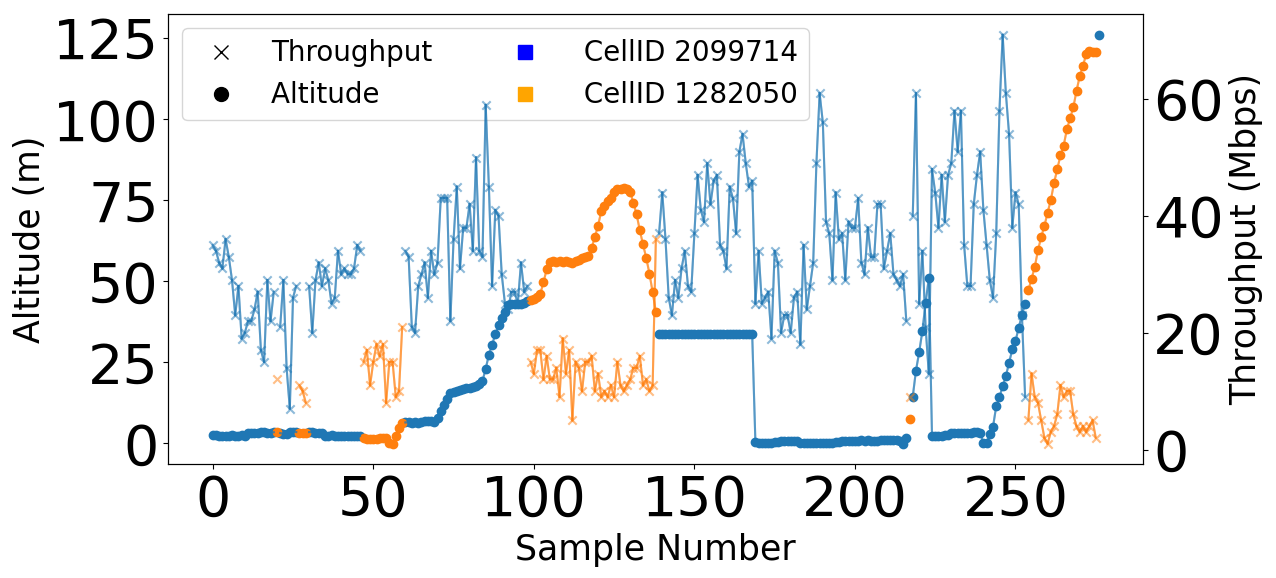}
    \vspace{-0.3cm}
    \caption{Handover compared to downlink throughput for trajectory 1.}
\vspace{-0.2cm}
    \label{fig:handover-alt}

\end{figure}

\begin{figure}[ht]
      \centering \includegraphics[scale=0.48]{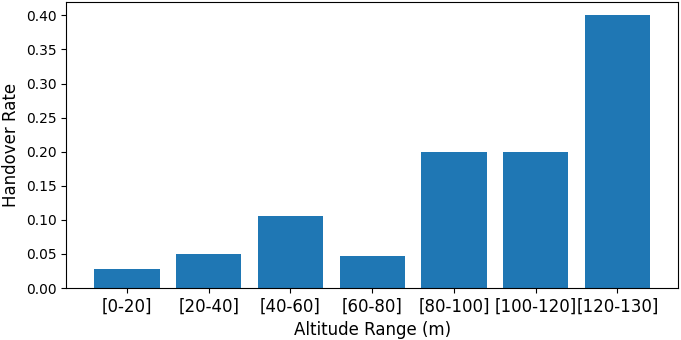}
      \vspace{-0.3cm}
    \caption{Handover rate in function of altitude range for trajectory 1.}
    \label{fig:handover-rate}
        \vspace{-0.3cm}
    \end{figure}

\begin{figure}[ht]
   \centering \includegraphics[scale=0.54]{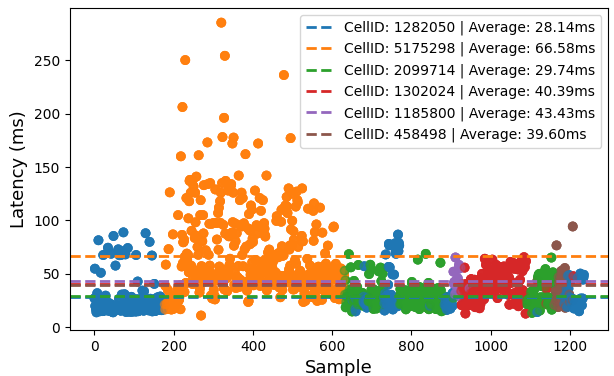}
   \vspace{-0.5cm}
    \caption{Handover compared to latency for trajectory 2.}
    \label{fig:latency}\vspace{-0.5cm}
\end{figure}

The network latency estimation process in our proposed architecture relies on real-world measurements to account for network variability. In particular, we concentrate on the non-reliable part of the network, which contributes to fluctuations in delay times. 
This unreliable segment is related to the link between the UAV and the edge server, where the elevated altitude and high speed of the UAV have a significant impact on handover and throughput. This impact is evident in the scenarios depicted in Figure \ref{fig:paths} and the measurements presented in Figures~\ref{fig:handover-alt}, \ref{fig:handover-rate}, and~\ref{fig:latency}. These figures illustrate the variations of handover measurement compared to throughput, altitude, and latency, respectively.

To obtain these measurements, we conducted a series of UAV flights, representing two scenarios. 
In the first scenario, the UAV maintained a fixed position while its altitude varied between $0-130 m$ as depicted in Figure \ref{fig:paths} (1) of the \textit{altitude trajectory}. 
In the second scenario, the UAV maintained a fixed altitude of $10 m$ and performed a linear back-and-forth trajectory of $100 m$ at a maximum speed of $5 m/s$ as shown in Figure \ref{fig:paths} (2) of the \textit{horizontal trajectory}.
During these UAV flights, UAV position, throughput, cell ID, speed, and NL were recorded.
This allowed us to construct the measurement database and closely examine how changes in altitude and position within a small area influenced the network's overall performance. By incorporating these real-world measurements into the DT, we can accurately estimate delays and make more informed decisions when relaying user commands to the UAV.

\section{Results and Analysis} 
\label{sec:results}

We present the results of two experimental scenarios depicted in Figure \ref{fig:paths}. 
Figures \ref{fig:handover-alt} and \ref{fig:handover-rate} depict two critical performance metrics for a UAV communication system during a handover between two base stations as altitude changes. Figure \ref{fig:handover-alt} shows the measured throughput as a function of the BS's cell IDs during the handover, while Figure \ref{fig:handover-rate} displays the handover rate as a function of altitude. Both figures relate to the first scenario shown in Figure \ref{fig:paths} (1), which illustrates the UAV's altitude  trajectory. Our observations indicate a considerable reduction in throughput from an average of $60 Mbps$ to $10 Mbps$ when the UAV ascends beyond $20 m$ altitude due to high interference from neighboring base stations in line of sight (LoS) with the UAV, which leads to increased handover rates. Figure \ref{fig:latency} presents latency versus handover data from the second scenario shown in Figure \ref{fig:paths} (2). We observed 6 \acp{bs} attachment with an increase in handover frequency near the lake area, representing the edges of different cells, while the average delay remained relatively constant with the distances to non-LoS \acp{bs}.

\begin{figure}[ht]
    \centering \includegraphics[scale=0.65]{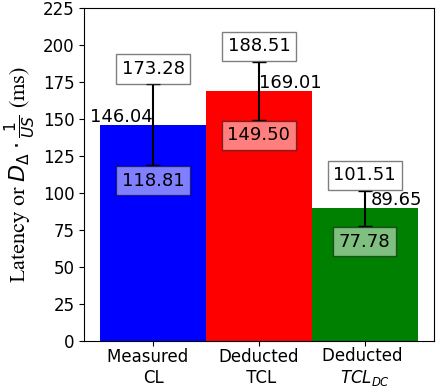}
    \vspace{-0.2cm}
    \caption{Measured latency and corresponding latency deducted from distances.}
    \label{fig:result}
\end{figure}

To validate our system and demonstrate the effectiveness of the DC method, we first measured the average End-to-End (E2E) CL between the DT and the real UAV. This measured CL showed an average delay of $146.04 ms$ with a standard deviation of $27.23 ms$.
We then conducted 20 UAV flight experiments with a VR operator sending a command flow to both the DT and the real UAV that was moving towards an obstacle with a speed ranging from $1-5 m/s$, without employing delay compensation. Theoretically, the difference between the traveled distance of the real UAV and the digital one, denoted as $D_{\Delta}$, would be proportional to the theoretical CL (TCL) described by Equation~(\ref{eq:theory}) as
\begin{equation}\label{eq:theory}
\small
D_{\Delta}=(CL + NL)\cdot US \implies TCL = D_{\Delta}\cdot \frac{1}{US}.
\end{equation}
The average TCL measured was $169.01 ms$ with a standard deviation of $19.5 ms$, highlighting the effectiveness of our DT physics in accurately mimicking the real world, as otherwise there would have been a significant difference between the two distances due to the inertia and frictional force with the air. 
The difference in the traveled distances matches the measured delay, with an average error of $22.96 ms$ ($14.57 \%$) due to the NL variation, demonstrating the reliability of our measurements and the realism of our implementation, thanks to considering weather variations and other real-world parameters.
Upon applying our proposed DC method, the TCL ($TCL_{DC}$) showed a significant reduction, with an average delay of $89.64 ms$ and a standard deviation of $11.86 ms$, achieving a delay reduction of $53.04 \%$ compared to the previous measurements. 
Figure \ref{fig:result} shows the comparison of the measured CL, deducted TCL, and deducted $TCL_{DC}$. 

\vspace{-0.3cm}

\section{Conclusion}
\label{sec:conclusion}

In conclusion, this paper presents a novel architecture that significantly enhances the reliability of UAV teleoperation in critical scenarios. By leveraging DT technology, we create an immersive virtual environment mirroring the physical world, incorporating 3D surroundings, weather constraints, and network limitations. Our architecture employs an intelligent logic that uses data from both the virtual and physical worlds to make informed decisions regarding operator-initiated actions. Furthermore, the proposed delay compensation method has shown significant improvement in reducing network latency, which makes this architecture even more promising for real-world deployment.  Field trials have substantiated the efficacy of this architecture, showcasing its potential in mitigating risks and enhancing UAV teleoperation reliability.

Looking ahead, the principles and mechanisms underscored herein can find meaningful integration within healthcare, facilitating remote medical interventions and surveillance. Furthermore, as the industry propels towards Industry 5.0, our architecture can play an instrumental role in fostering seamless human-machine collaboration, driving operational efficiencies and ushering in a new era of industrial symbiosis. Future research can explore the integration of advanced artificial intelligence algorithms for decision-making, and autonomous flight capabilities.

\printbibliography

@article{malik2020effect,
  title={Effect of haptic feedback on pilot/operator performance during flight simulation},
  author={Malik, Hassam Ahmed and Rasool, Shahzad and Maqsood, Adnan and Riaz, Rizwan},
  journal={Applied Sciences},
  volume={10},
  number={11},
  pages={3877},
  year={2020},
  publisher={MDPI}
}

@inproceedings{hoppe2018vrhapticdrones,
  title={VRHapticDrones: Providing haptics in virtual reality through quadcopters},
  author={Hoppe, Matthias and Knierim, Pascal and Kosch, Thomas and Funk, Markus and Futami, Lauren and Schneegass, Stefan and Henze, Niels and Schmidt, Albrecht and Machulla, Tonja},
  booktitle={Proceedings of the 17th International Conference on Mobile and Ubiquitous Multimedia},
  pages={7--18},
  year={2018}
}

@inproceedings{abtahi2019beyond,
  title={Beyond the force: Using quadcopters to appropriate objects and the environment for haptics in virtual reality},
  author={Abtahi, Parastoo and Landry, Benoit and Yang, Jackie and Pavone, Marco and Follmer, Sean and Landay, James A},
  booktitle={Proceedings of the 2019 CHI Conference on Human Factors in Computing Systems},
  pages={1--13},
  year={2019}
}

@article{liu2021review,
  title={Review of digital twin about concepts, technologies, and industrial applications},
  author={Liu, Mengnan and Fang, Shuiliang and Dong, Huiyue and Xu, Cunzhi},
  journal={Journal of Manufacturing Systems},
  volume={58},
  pages={346--361},
  year={2021},
  publisher={Elsevier}
}

@article{rosen2015importance,
  title={About the importance of autonomy and digital twins for the future of manufacturing},
  author={Rosen, Roland and Von Wichert, Georg and Lo, George and Bettenhausen, Kurt D},
  journal={Ifac-Papersonline},
  volume={48},
  number={3},
  pages={567--572},
  year={2015},
  publisher={Elsevier}
}

@article{tu2021mixed,
  title={A mixed reality interface for a digital twin based crane},
  author={Tu, Xinyi and Autiosalo, Juuso and Jadid, Adnane and Tammi, Kari and Klinker, Gudrun},
  journal={Applied Sciences},
  volume={11},
  number={20},
  pages={9480},
  year={2021},
  publisher={MDPI}
}

@inproceedings{concannon2019quality,
  title={A quality of experience evaluation system and research challenges for networked virtual reality-based teleoperation applications},
  author={Concannon, David and Flynn, Ronan and Murray, Niall},
  booktitle={Proceedings of the 11th ACM workshop on immersive mixed and virtual environment systems},
  pages={10--12},
  year={2019}
}

@article{tao2022digital,
  title={Digital twin modeling},
  author={Tao, Fei and Xiao, Bin and Qi, Qinglin and Cheng, Jiangfeng and Ji, Ping},
  journal={Journal of Manufacturing Systems},
  volume={64},
  pages={372--389},
  year={2022},
  publisher={Elsevier}
}

@article{10.3389/frvir.2023.1019080,
  title={TwinXR: Method for using digital twin descriptions in industrial eXtended reality applications},
  author={Tu, Xinyi and Autiosalo, Juuso and Ala-Laurinaho, Riku and Yang, Chao and Salminen, Pauli and Tammi, Kari},
  journal={Frontiers in Virtual Reality},
  volume={4},
  pages={1019080},
  year={2023},
  publisher={Frontiers}
}

@article{sun2020reducing,
  title={Reducing offloading latency for digital twin edge networks in 6G},
  author={Sun, Wen and Zhang, Haibin and Wang, Rong and Zhang, Yan},
  journal={IEEE Transactions on Vehicular Technology},
  volume={69},
  number={10},
  pages={12240--12251},
  year={2020},
  publisher={IEEE}
}

@article{kikuchi2022future,
  title={Future landscape visualization using a city digital twin: Integration of augmented reality and drones with implementation of 3D model-based occlusion handling},
  author={Kikuchi, Naoki and Fukuda, Tomohiro and Yabuki, Nobuyoshi},
  journal={Journal of Computational Design and Engineering},
  volume={9},
  number={2},
  pages={837--856},
  year={2022},
  publisher={Oxford University Press}
}

@article{yang2022extended,
  title={Extended Reality Application Framework for a Digital-Twin-Based Smart Crane},
  author={Yang, Chao and Tu, Xinyi and Autiosalo, Juuso and Ala-Laurinaho, Riku and Mattila, Joel and Salminen, Pauli and Tammi, Kari},
  journal={Applied Sciences},
  volume={12},
  number={12},
  pages={6030},
  year={2022},
  publisher={MDPI}
}

@article{kim2021p300,
  title={P300 brain-computer interface-based drone control in virtual and augmented reality},
  author={Kim, Soram and Lee, Seungyun and Kang, Hyunsuk and Kim, Sion and Ahn, Minkyu},
  journal={Sensors},
  volume={21},
  number={17},
  pages={5765},
  year={2021},
  publisher={MDPI}
  }

@article{9951155,
  title={VR-based immersive service management in B5G mobile systems: A UAV command and control use case},
  author={Taleb, Tarik and Sehad, Nassim and Nadir, Zinelaabidine and Song, Jaeseung},
  journal={IEEE Internet of Things Journal},
  volume={10},
  number={6},
  pages={5349--5363},
  year={2022},
  publisher={IEEE}
}

@inproceedings{hayat2019experimental,
  title={An experimental evaluation of LTE-A throughput for drones},
  author={Hayat, Samira and Bettstetter, Christian and Fakhreddine, Aymen and Muzaffar, Raheeb and Emini, Driton},
  booktitle={Proceedings of the 5th Workshop on Micro Aerial Vehicle Networks, Systems, and Applications},
  pages={3--8},
  year={2019}
}

@article{michel2004cyberbotics,
  title={Cyberbotics ltd. webots™: professional mobile robot simulation},
  author={Michel, Olivier},
  journal={International Journal of Advanced Robotic Systems},
  volume={1},
  number={1},
  pages={5},
  year={2004},
  publisher={SAGE Publications Sage UK: London, England}
}

@article{alsamhi2022computing,
  title={Computing in the sky: A survey on intelligent ubiquitous computing for uav-assisted 6g networks and industry 4.0/5.0},
  author={Alsamhi, Saeed Hamood and Shvetsov, Alexey V and Kumar, Santosh and Hassan, Jahan and Alhartomi, Mohammed A and Shvetsova, Svetlana V and Sahal, Radhya and Hawbani, Ammar},
  journal={Drones},
  volume={6},
  number={7},
  pages={177},
  year={2022},
  publisher={MDPI}
}

@article{10071541,
  title={Locomotion-Based UAV Control Towards the Internet of Senses},
  author={Sehad, Nassim and Cherif, Bilel and Khadraoui, Ibrahim and Hamidouche, Wassim and Bader, Faouzi and J{\"a}ntti, Riku and Debbah, M{\'e}rouane},
  journal={IEEE Transactions on Circuits and Systems II: Express Briefs},
  year={2023},
  publisher={IEEE},
  year={2023},
  volume={70},
  number={5},
  pages={1804-1808},
}

\end{document}